\documentclass[onecolumn]{aastex62}
\usepackage{graphicx}

\graphicspath{{images/}}

\usepackage{booktabs}
\usepackage{amsmath}
\usepackage{graphicx}

\usepackage{epsfig}
\usepackage{natbib}
\usepackage{wasysym}

\usepackage{siunitx}
\usepackage{lipsum}
\usepackage{mathrsfs}
\usepackage{float}
\usepackage{rotating}
\usepackage{epstopdf}
\usepackage[graphicx]{realboxes}
\usepackage{array}
\newcolumntype{P}[1]{>{\centering\arraybackslash}p{#1}}
\newcolumntype{M}[1]{>{\centering\arraybackslash}m{#1}}
\newcounter{chem}
\newcounter{temp}
\usepackage{lineno}
\usepackage{sidecap}
\usepackage{colortbl}
\usepackage{tabularx}

\usepackage{color}

\usepackage{gensymb}

\usepackage{filecontents}
\usepackage{filecontents}

\usepackage{lineno}

\begin{document}

\shorttitle{Topography \& Climate Transition}
\shortauthors{Chen, Ildirimzade, \& Macdonald}

\title{{\bf Topography-Induced Stationary Waves and the Onset of Nightside Warming on Rocky Planets around M-dwarf Stars}}

\author[0000-0003-1995-1351]{Howard Chen}

\affil{Department of Aerospace, Physics, and Space Sciences, Florida Institute of Technology, Melbourne, FL}

\affil{Sellers Exoplanet Environments Collaboration (SEEC), NASA Goddard Space Flight Center,  USA}

\author{Aida Ildirimzade}

\affil{Department of Aerospace, Physics, and Space Sciences, Florida Institute of Technology, Melbourne, FL}

\author[0000-0001-5540-3817]{Evelyn Macdonald}
\affil{Department of Astrophysics, Faculty of Earth Sciences, Geography and Astronomy, University of Vienna, Vienna, Austria}

\begin{abstract}
Rocky planets orbiting within the habitable zones of M dwarfs are one of the most accessible planet populations for atmospheric characterization, yet their climates may differ substantially from those of Earth analogs. In the tidally locked limit, the nightside's tendency to radiatively cool and potentially trap volatiles as permanent ice introduces a strong dependence of habitability on the planet's surface and atmospheric boundary conditions. We perform a suite of synchronously rotating exoplanet climate experiments spanning a series of topographic and orographic realizations with different mean elevations and landmass distributions. Across a grid of $p_{\mathrm{N2}} = 0.5-8~\mathrm{bar}$ and $F_{\oplus} = 1200-1700~\mathrm{W\,m^{-2}}$, we find that surface relief breaks the flow symmetry, replacing the circumpolar vortices with mechanically forced stationary waves. For example, steep orography produces standing Rossby gyres that strengthen the cross-terminator jet and align vertical uplift with the day--night boundary. These new circulation regimes enhance moisture transport, increasing the infrared optical depth and promoting additional nightside cloud formation, which produces a stronger cloud-greenhouse feedback. Broad, elevated plateaus drive a similarly fragmented but slightly weaker circulation, yielding less effective moisture transport. These results show that the relief and spatial distribution of landmasses, parameters unconstrained for most exoplanets, can exert strong controls on the climatic bifurcations of tidally locked M-dwarf exoplanets.
\end{abstract}


\section{Introduction}

Rocky planets orbiting M dwarfs now dominate the known inventory of potentially habitable worlds.  Their small host stars provide favorable planet-to-star contrast ratios and short orbital periods, enabling the detection of terrestrial planets within the nominal habitable zone through both transit and radial-velocity techniques \citep{dressing2015occurrence}.  Systems such as TRAPPIST-1, LHS~1140, and TOI~715  suggest that compact architectures of Earth-sized planets around late-type stars are common \citep{GillonEt2017,ment2019,dransfield}. Despite the strong activity of M-dwarfs, their planets hold some promise as potential laboratories for comparative exoplanet climatology and the search for temperate, habitable atmospheres.

The climate of tidally locked M-dwarf planets, however, departs drastically from Earth analogs.  Owing to synchronous rotation, one hemisphere receives constant stellar irradiation while the other remains in perpetual darkness.  This geometry drives a  dichotomy between day–night energy transport and radiative cooling \citep{joshi2003climate,merlis2010,turbet2016}.  When atmospheric pressure or greenhouse trapping are insufficient, the nightside can cool to the point where volatiles condense irreversibly, producing a ``snowball'' or ``eyeball'' state characterized by a small open-water region near the substellar point \citep{pierrehumbert2010}.  Increasing background gas pressure or stellar flux can trigger a sharp transition to a ``temperate–nightside'' regime, where enhanced advection and latent heating maintain above-freezing temperatures across a larger fraction of the surface \citep{yang2013stabilizing,Macdonald2025}.  

\begin{figure*}[h!]
    \centering
    \includegraphics[width=1.0\textwidth]{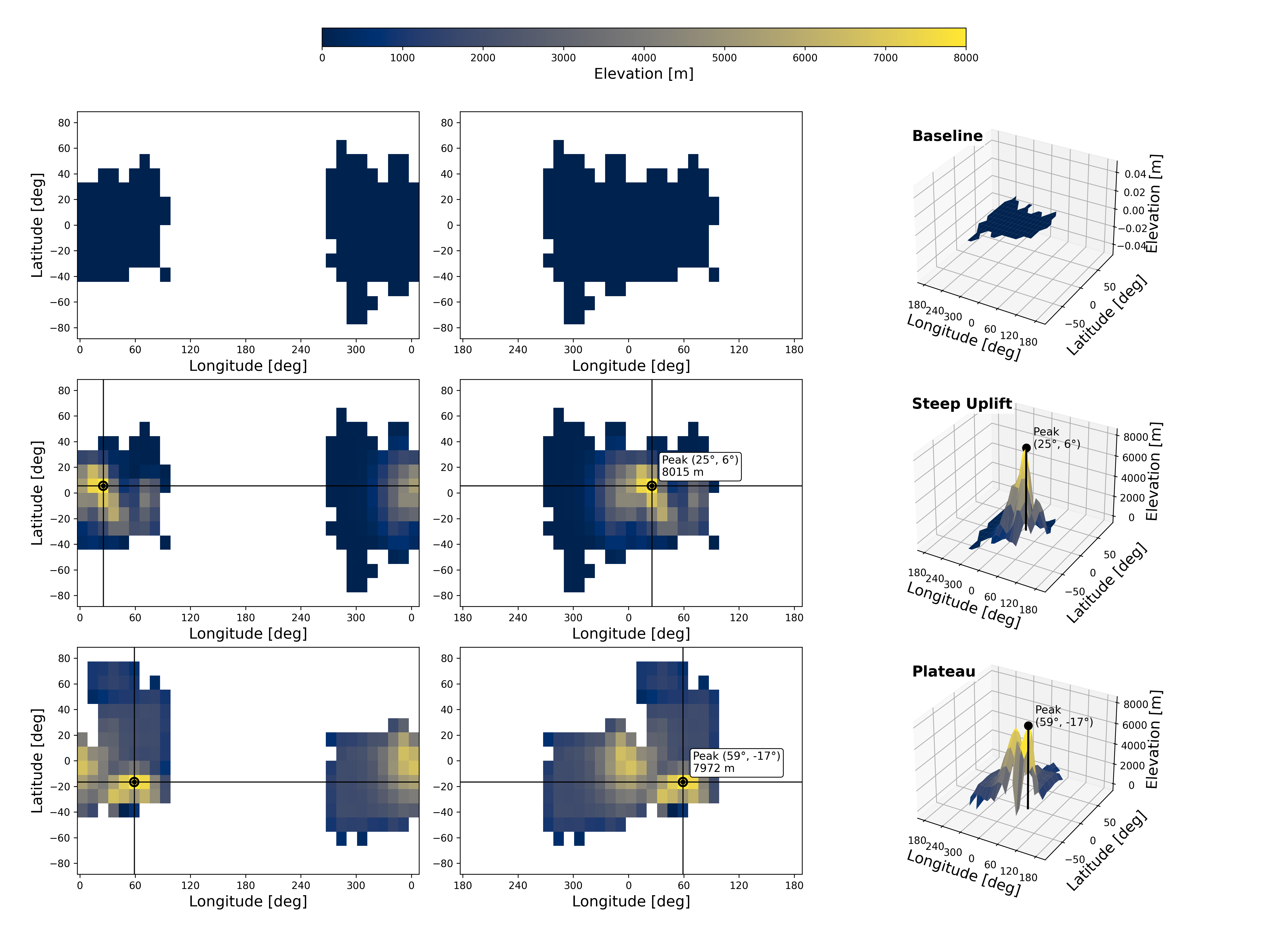}
    \caption{ Surface boundary configurations used in this study.  {\it Middle column:} map view of continental distribution and topography with the substellar point centered at 0° longitude, or on the dayside.  {\it Left column:} identical surface realizations centered on the nightside. {\it Right column:} corresponding three-dimensional renderings of the topography. Rows show (top) Baseline flat continent, (middle) Steep Uplift, and (bottom) Dayside Plateau. Although the mean elevation is comparable in the bottom two cases, the spatial distribution of relief differs substantially and sets up distinct stationary-waves and cross-terminator flow in the simulations.}
    \label{fig:topography}
\end{figure*}

Despite rapid progress in modeling these transitions, nearly all previous studies have adopted idealized aquaplanet, uniform-land surfaces, or Earth-based continental configurations and ocean bathymetries (e.g., \citealt{ChenEt2019ApJ,chen2018apjl}). Previous work has shown that the size and location of landmasses have crucial consequences on the atmospheric circulation and chemistry of tidally locked exoplanets \citep{Lewis+2018,SalazarEt2020ApJL,macdonald2022}, for both ocean-rich Earth-like planets \citep{sainsbury2024} and arid planets \citep{glaser2025}; a diversity that is expected due to the wide range of possible volatile accretion invetories and histories \citep{suer2023,bower2025,chen2025}. However, the influence of continental geometry and topography remains largely unexplored.  On Earth, elevated Plateaus and asymmetric land–sea distributions exert major control over monsoon systems, jet placement, and regional temperature contrasts \citep{boos2021,kaspi2015}.  For tidally locked planets, such effects may be amplified: substellar highlands could enhance local heating and suppress low-level advection, while nightside mountain ranges might favor cold-trap formation and volatile sequestration.  Yet to our knowledge, no existing study has numerically quantified how the amplitude and spatial distribution of topography shape the stability of these climate regimes.

Another important reason to examine the effects of topography on climate is that the long-term habitability of rocky exoplanets is widely theorized to depend on the carbonate-silicate weathering thermostat, a cyclical process driven by active plate tectonics and volatile recycling (e.g., \citealt{KastingEt1993Icarus,mcintyre2022,stamenkovic2016}). An important  consequence of this tectonic activity is orogeny, or mountain building. Therefore, if a temperate rocky exoplanet possesses the geophysical mechanisms required to regulate its climate and sustain life, it must inherently possess significant surface topography. 
 
 Earth-similar land–ocean distributions and orographic configurations provide a valuable reference point  (e.g., \citealt{bhongade2024,sainsbury2024}). However, recent geological perspectives suggest that rocky exoplanets will naturally develop diverse topographic relief through dynamic topography, petrologic evolution of continents, and impact cratering \citep{guimond2026}. These possibilities imply varied water-to-land ratios and have examined their consequences for planetary energy balance via surface albedo and the carbonate-silicate weathering thermostat. The implications of such topographic diversity for atmospheric dynamics have received less attention.  Here, we consider a range of idealized orographic regimes with slightly different landmass distributions. Our numerical simulations explore a range of  ($P_{N2}$, $F_*$, Topography) parameter space using a reduced complexity 3D general circulation model to demonstrate how variations in topography and land distribution drive departures from the canonical M-dwarf planetary climates.


\section{Numerical Model \& Experiments}

All simulations were performed using the intermediate-complexity general circulation model \texttt{ExoPlaSim} \citep{Paradise+22}. While high-complexity GCMs (e.g., ExoCAM, ROCKE-3D, \citealt{wolf2022,way2017}) offer greater spectral fidelity, their computational cost prohibits the broad parameter sweep. \texttt{ExoPlaSim} provides a necessary compromise, capturing essential moist dynamics and radiative drivers at a speed that enables a greater exploration of the ($P_{N2}$, $F_*$, Topography) phase space.  

The model couples a two-band shortwave and one-band longwave radiative transfer scheme with prognostic water vapor, sea ice, and moist convection following the configuration described by \citet{macdonald2024} and  \citet{Macdonald2025}\footnote{An initial subset of these simulations was presented in \citet{macdonald2024} , with additional cases and a more comprehensive analysis of the resulting climates reported in \citet{Macdonald2025}.}.  The host star is treated as a blackbody with an effective temperature of $2600~\mathrm{K}$, representative of a late-M dwarf.  Rayleigh scattering, surface albedo, and stellar spectral weighting are implemented following \citet{Paradise+22}.  All planets are synchronously rotating with planet radii of $0.9\,R_{\oplus}$ and  surface gravities of $7~\mathrm{m\,s^{-2}}$.
\begin{table}[t]
\centering
\caption{
For each surface configuration, we explored a common grid of atmospheric parameters following \citet{Macdonald2025}, spanning nitrogen partial pressures
$p_{\mathrm{N2}} = \{0.5, 2, 6, 8 \}$~bar and stellar fluxes
$F_{\star} = \{1200, 1400, 1600, 1700\}$~W\,m$^{-2}$.
The lowest-flux case approximates the outer edge of the temperate regime, while $1600-1700~\mathrm{W\,m^{-2}}$ approaches the inner edge where fully temperate-nightside climates first emerge.
All simulations assume synchronous rotation, identical planetary radius and resolution (T21; 32~$\times$~64), and atmospheres containing a fixed 1~mbar background of CO$_2$ with no ozone.
}
\label{tab:experiment_grid}
\begin{tabular}{lccc}
\toprule
\textbf{Surface Configuration} &
\textbf{Peak Elevation} &
\textbf{Gaussian Width} \\
\midrule
Baseline Substellar Continent & 0 km      & -- \\
Steep Uplift                 & $\sim$8 km & $20^{\circ}$ \\
Dayside Plateau              & $\sim$7 km & $30^{\circ}$ \\
\bottomrule
\end{tabular}
\end{table}

\subsection{Surface Boundary Conditions}

To isolate the climatic effects of topography and dayside land distribution, we constructed a hierarchy of three surface configurations (Table~\ref{tab:experiment_grid}). All land distributions were generated using the \texttt{randomcontinents.py} utility distributed with \texttt{ExoPlaSim}, ensuring identical planetary radius, grid resolution (T21; 32~$\times$~64), and synchronous rotation. Note that we adopt 8 km as a  upper limit reflecting the maximum isostatically supported relief observed across terrestrial bodies in the Solar System. Further, topography amplitudes larger than ${\sim}$1 scale height require a resolution jump that is computationally prohibitive for a parameter space study. 



Simulations were initialized from an isothermal $280~\mathrm{K}$ atmosphere with a uniform $30~\mathrm{m}$ mixed-layer ocean (where applicable) and integrated for 30~yr until the top-of-atmosphere (TOA) energy imbalance converged to less than $0.5~\mathrm{W\,m^{-2}}$.  Time-averaged diagnostics were computed over the final 10~yr of each integration.  All runs employed the T21 spectral resolution (32~$\times$~64) to permit multi-millennium equilibration for the densest atmospheres, consistent with \citet{Macdonald2025}. We set the substellar point to be at 0$\degree$ longitude in all cases.


\subsection{Numerical Validation and Sensitivity Tests}

We performed two classes of sensitivity tests following the methodology of \citet{Macdonald2025}.  

\paragraph{Advection suppression.}  We introduced an artificial Rayleigh drag term that damps horizontal winds with an exponential timescale $\tau_{\mathrm{drag}}$.  For a representative SubPlateau planet with $p_{\mathrm{N2}}=8~\mathrm{bar}$ and $F_{\star}=1600~\mathrm{W\,m^{-2}}$, the control run ($\tau_{\mathrm{drag}} \rightarrow \infty$) was in the temperate-nightside regime, whereas $\tau_{\mathrm{drag}}=10$~days produced a transitional climate and $\tau_{\mathrm{drag}}=1$~day recovered an eyeball state, demonstrating that efficient advection is essential for sustaining a warm nightside.

\paragraph{Dry-atmosphere experiments.}  For the same configuration, we disabled surface evaporation to remove water vapor feedbacks.  The resulting climate reverted to an eyeball state with a frozen nightside and a narrow dayside hot spot, showing that water vapor transport, latent heating, and its radiative effects drive the deglaciation transition.

\begin{figure*}[t!]
    \centering
    \includegraphics[width=1.05\columnwidth]{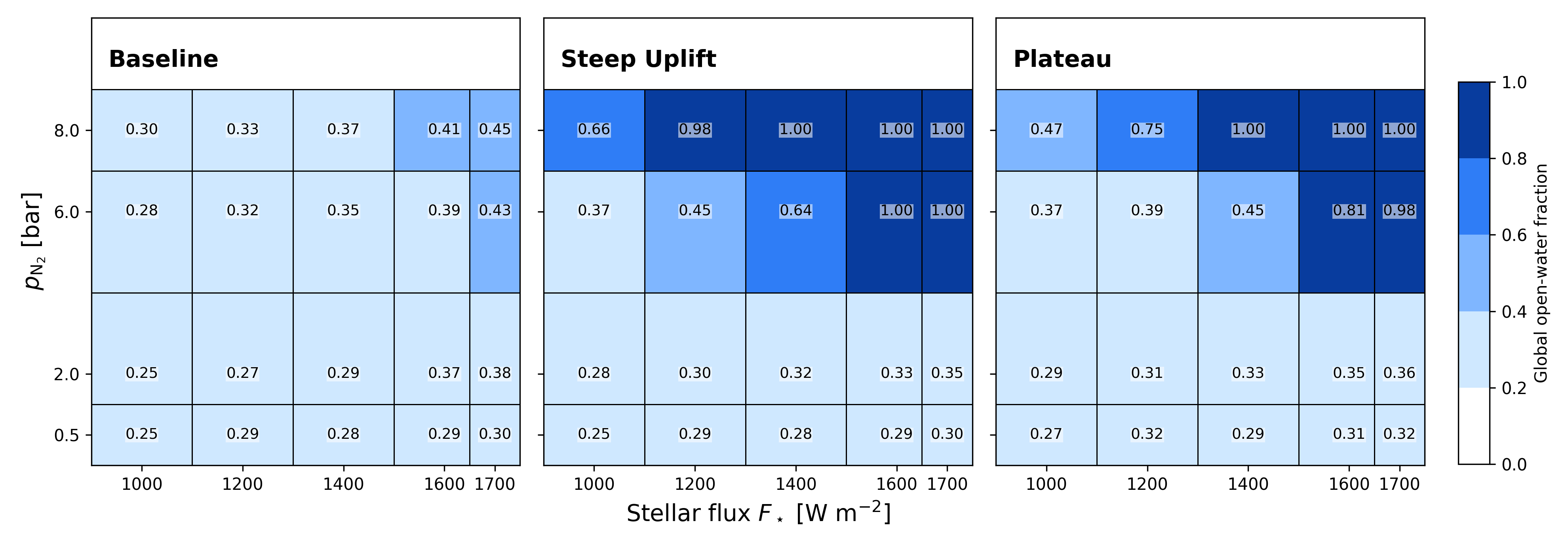}
    \caption{Open-water fraction across the $(F_\star, p_{\mathrm{N_2}})$ parameter grid. Color indicates the global fractional area that remains ice-free at equilibrium, with darker
    shades marking more extensive open-water regions. At low incident flux ($1200$--$1400~\mathrm{W\,m^{-2}}$), all
    simulations remain predominantly glaciated. At $1600~\mathrm{W\,m^{-2}}$, atmospheres exceeding a few bars develop
    measurable open-water areas, marking the onset of transitional climates. These maps summarize the broader parameter
    sweep referenced in Sections~3.1--3.2 and provide the context for the detailed single-forcing analyses presented
    later in the manuscript.}
    \label{fig:sum}
\end{figure*}

\section{Results}
The climatic response of synchronously rotating planets to variations in surface relief include: (1) the modification of stationary‐waves that regulates horizontal and vertical circulation, and (2) the resulting redistribution of moisture, clouds, and longwave energy between the day and nightside hemispheres.  In what follows, we examine how these mechanisms evolve across three representative topographic configurations, the Baseline Substellar Continent, Steep Uplift, and Dayside Plateau, We first present an overview of climate states across the parameter grid, then focus on a representative transitional regime ( $p_{\mathrm{N2}} = 6~\mathrm{bar}$ and $F_{\star} = 1600~\mathrm{W\,m^{-2}}$) to elucidate the dynamical mechanism.  This regime resides near the transition between nightside freezing and fully temperate conditions, making it ideal for isolating dynamical feedbacks that govern cross‐terminator energy exchange.  The subsequent subsections progress from the global circulation and stationary‐wave structure to their thermodynamic consequences for clouds, longwave fluxes, and nightside temperature.

The three principal surface boundary configurations used in this study are shown in Figure~\ref{fig:topography}: a Baseline substellar Plateau, Steep Uplift chains, and an elevated but spatially extended Plateau. 
Although all share comparable mean elevations ($\sim$5--8~km), their spatial distributions produce markedly different dynamical responses.
These distinctions establish the dynamical backdrop for the thermodynamic contrasts described below.

\subsection{Topographic Induced Climatic Influence}

\begin{figure*}[t!]
    \centering
    \includegraphics[width=\textwidth]{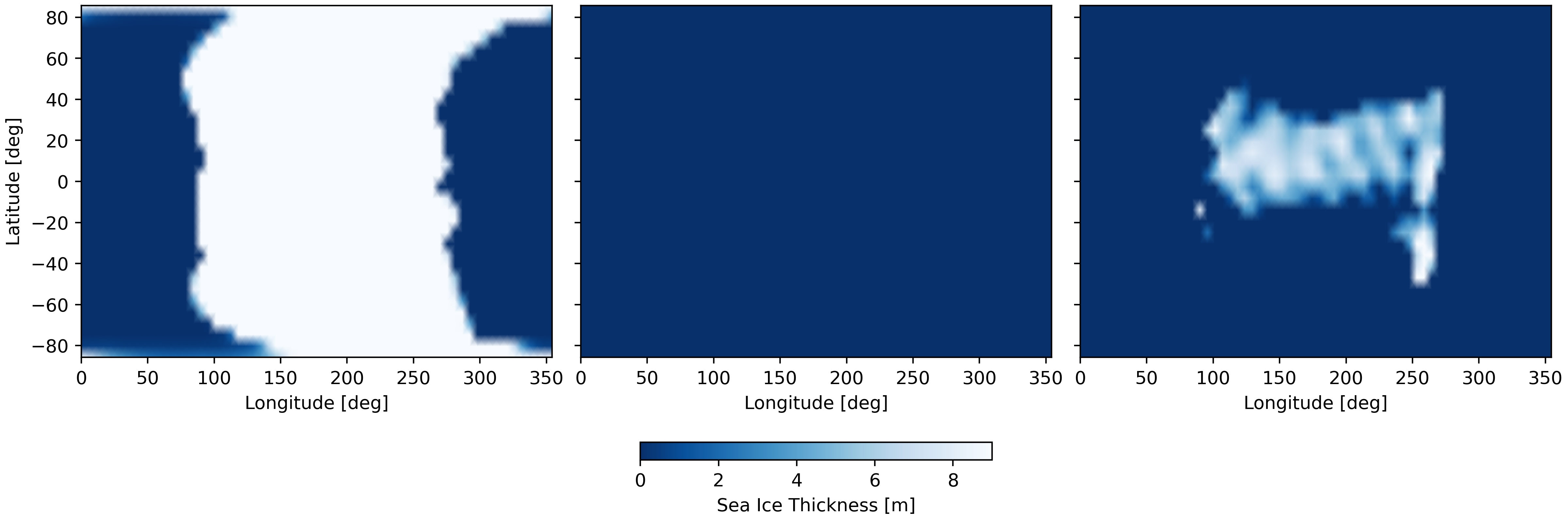}
    \caption{\textbf{Equilibrium sea-ice thickness for the three topographies.} The panels shown are for:
    (\textit{Left}) Baseline, (\textit{center}) Steep Uplift, and (\textit{right}) Plateau.
    Topographic control of the circulation strongly modulates volatile sequestration. 
    The Steep Uplift case lead to open nightside ocean, while the Plateau tends reduced nightside ice and patchy thin ice emcompassing the antistellar point.
    Gray outlines mark the sea ice versus open water divide. Maximum sea ice is set at 9 m in the model.}
    \label{fig:seaice}
\end{figure*}

Figure~\ref{fig:sum} provides an overview of the simulated climate states across the $(F_\star, p_{\mathrm{N_2}})$ parameter grid for the Baseline, Steep Uplift, and Dayside Plateau configurations.  At low incident flux ($F_\star = 1200$--$1400~\mathrm{W\,m^{-2}}$) and low pressures ($p_{\mathrm{N_2}} \le 2.0$~bar), the climate state is relatively insensitive to surface relief; all three configurations remain in a predominantly glaciated ``eyeball'' state with open-water fractions confined to $\sim0.28$--0.35.  

In general higher $p_{\mathrm{N_2}}$ ($\gtrsim 6$--8~bar) leads to greater longwave absorption which is amplified by pressure broadening, yielding   warmer, less glaciated states. Further in this regime, the climatic response diverges sharply between the surface types. While the Baseline case exhibits only a gradual expansion of the open ocean, reaching a maximum fraction of 0.46 at the highest flux, the topographic configurations undergo a rapid transition to fully temperate conditions. For instance, at $F_\star = 1400~\mathrm{W\,m^{-2}}$ and 8.0~bar, the Steep Uplift and Plateau cases reach unity (1.00) open-water fraction, representing global deglaciation, whereas the Baseline remains at 0.37. This contrast is particularly distinct at $1600~\mathrm{W\,m^{-2}}$ and 6.0~bar, where the Steep Uplift planet is fully ice-free (1.00) and the Plateau is nearly so (0.81), yet the Baseline case retains extensive nightside ice with an open-water fraction of only 0.39. These diagnostics establish the broader context for the single-forcing cases examined in
Sections~3.1--3.2 and delineate the regions of parameter space in which fully glaciated, transitional, and partially temperate climates emerge.

Figure~\ref{fig:seaice} illustrates the equilibrium sea-ice distribution for the three topographies (Baseline Substellar Continent, Steep Uplift, and Dayside Plateau) at a flux of 
$F_\star = 1600~\mathrm{W\,m^{-2}}$ and $p_{\mathrm{N_2}} = 6~\mathrm{bar}$.
The Baseline planet maintains a classical ``eyeball'' configuration---open water near the substellar point surrounded by a thick nightside ice shell.
Introducing steep mountains dramatically reduces nightside ice coverage: topography enhances cross-terminator advection and leads to complete removal of sea-ice.
In contrast, the Plateau experiment exhibits thin, patchy ice out to $\sim90^\circ$ longitude similar to the low-$p_{\mathrm{N_2}}$ Baseline runs.

\begin{figure*}[t!]
    \centering
    \includegraphics[width=\textwidth]{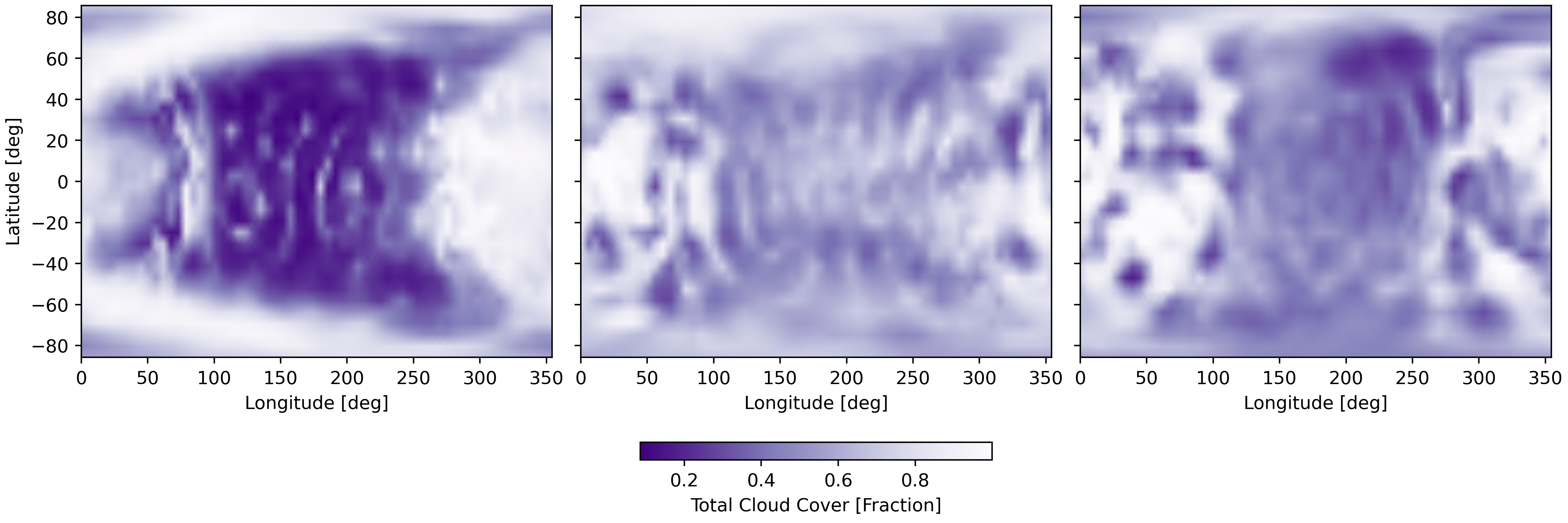}
    \caption{\textbf{Time-mean total cloud cover.} The panels shown are for:
    (\textit{Left}) Baseline, (\textit{center}) Steep Uplift, and (\textit{right}) Plateau.
    The Baseline case exhibits the conventional tidally locked cloud distribution, with enhanced cloudiness on the dayside and reduced cloudiness on the nightside. Orography enhances ascent and cloudiness. The Steep case produces the highest cloud cover at the nightside compared to the other two cases. The Plateau case yields intermediate nightside cloud cover, with cloudiness distributed along the day-night terminators.}
    \label{fig:clouds}
\end{figure*}

Total cloud cover (Figure~\ref{fig:clouds}) varies substantially across our simulations, revealing a strong dependence on the topographic scale. The Baseline case adheres to the conventional tidally locked configuration, showing the lowest total cloud fraction with maxima confined to the dayside and minimal cloudiness on the nightside. This pattern is greatly modified by the introduction of topography. The Steep Uplift case produces the highest cloud fraction overall, characterized by distinctly high cloud  the nightside hemisphere. This concentration of nightside cloudiness acts as a potent infrared blanket, significantly enhancing the nightside greenhouse effect relative to the other configurations. The Plateau case represents an intermediate regime, generating greater total cloud cover than Baseline but distributing it primarily along the day–night terminators.

\begin{figure*}[t!]
    \centering
    \includegraphics[width=\textwidth]{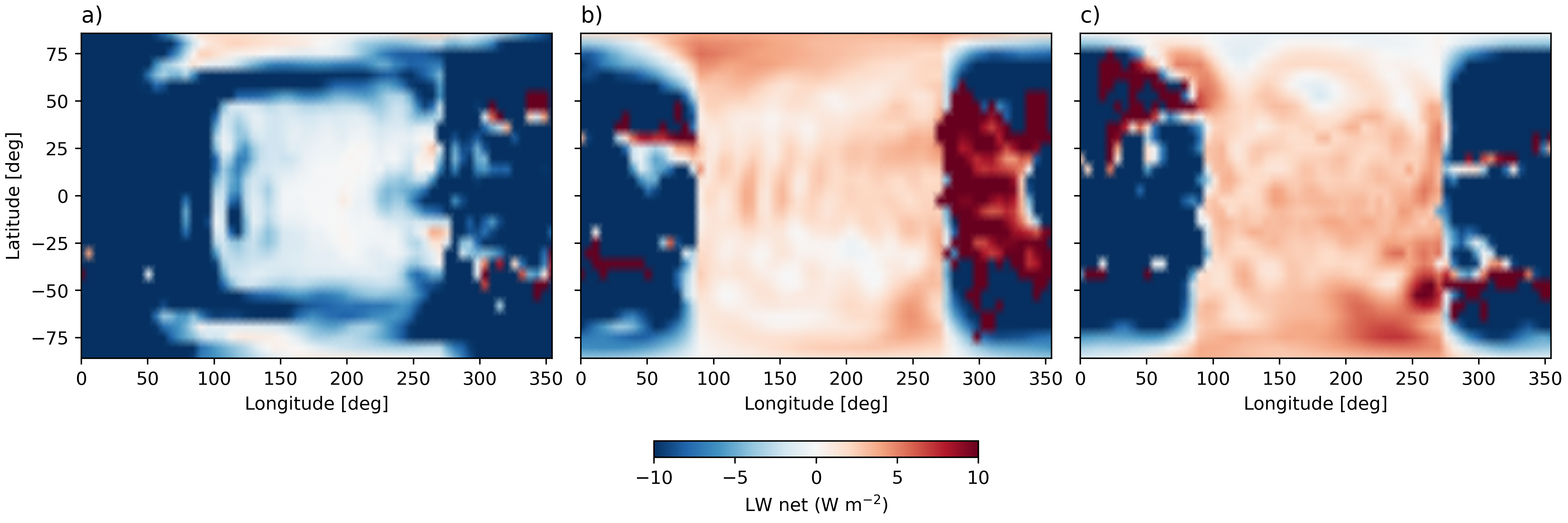}
    \caption{{\bf Net longwave flux at the planetary surface.} Panels: (a) Baseline, (b) Steep Uplift, (c) Plateau. Topography enhances nightside cloud cover, increasing downwelling longwave flux and warming the surface. Unlike the clear‑sky Baseline, both the Steep Uplift and Plateau cases exhibit broader warming centered near the anti‑stellar point. The Steep Uplift case alone shows a distinct “red‑tongue” warming feature, driven by a stationary Rossby wave that channels moist air eastward.l.}
    \label{fig:longwave}
\end{figure*}

From the map projections of net LW flux at the surface (Figure~\ref{fig:longwave}), both topographic cases exhibit broader warming (redder colors) centered on the anti-stellar point ($180^{\circ}$), which corresponds to the region of maximum cloud cover. Unlike the Baseline, where the deep nightside remains relatively clear and radiatively open to space, the topography enhances moisture advection into the anti-stellar region, sustaining a widespread, optically thick cloud deck. These clouds exert a positive surface longwave cloud radiative effect by significantly increasing downwelling thermal emission. This increased cloud formation creates the observed positive net LW flux anomaly, effectively insulating the nightside surface against radiative cooling. 

While the Baseline surface cooling is dominated by simple land-sea thermal contrast, the Steep Uplift case reveals a distinct downstream heating feature. This 'red tongue' is driven by a topography-induced stationary Rossby wave, which channels a concentrated plume of moisture eastward. The high specific humidity and adiabatic warming within this plume generate a strong local greenhouse effect that enhances downwelling radiation, helping to maintain a liquid ocean. The Plateau case lacks this intense, localized feature because its shallow gradient is insufficient to mechanically force a coherent stationary wave train.

\begin{figure*}[t!]
    \centering
    \includegraphics[width=\textwidth]{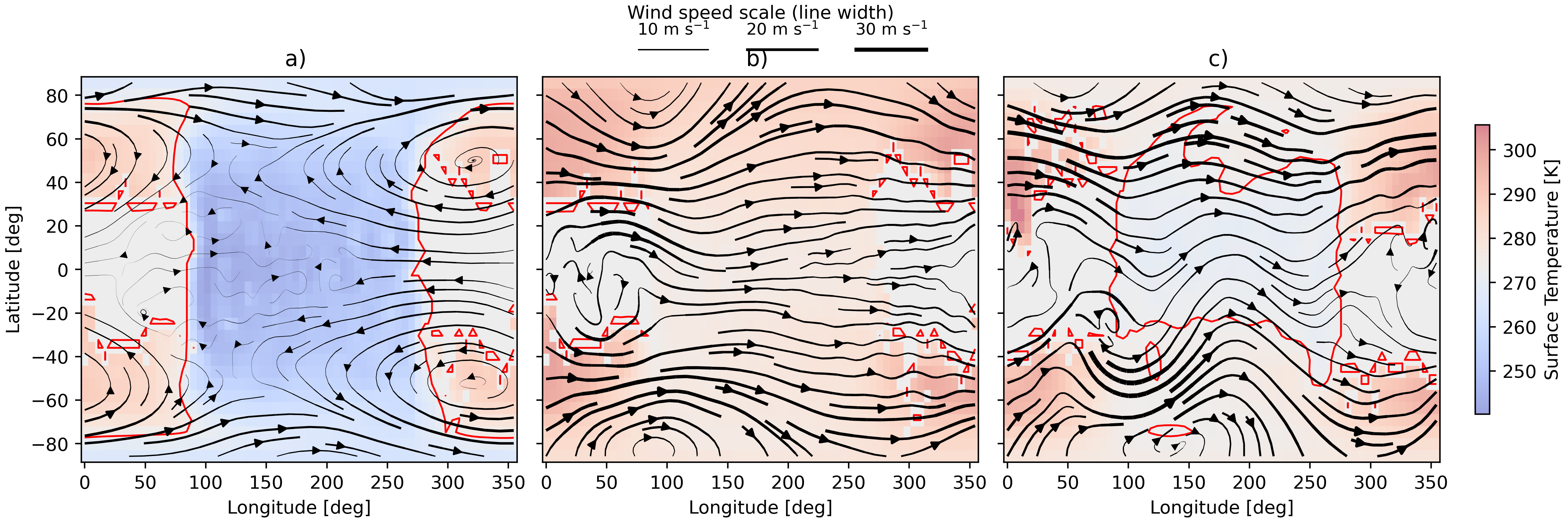}
    \caption{\textbf{Global wind velocity streamlines for the three topographic configurations.}
    Black streamlines show time-mean horizontal flow at 100~hPa, with line width proportional to wind speed (see scale bar).
    Red contours outline the 273 K temperature curve. The panels shown are for:
    (\textit{Left}) Baseline, (\textit{center}) Steep Uplift, and (\textit{right}) Plateau.
    In this regime, the Steep Uplift and Plateau cases exhibit intensified stationary waves and cross-terminator flow, while the Baseline yields a more zonally symmetric and weaker circulation.}
    \label{fig:streamlines}
\end{figure*}

Examining the global circulation patterns across the three topographic cases (Figure~\ref{fig:streamlines}), we find that the main difference between the topographic cases and the baseline is that the near-surface winds become more convergent toward the substellar highlands, feeding localized updrafts and cloud formation. 
In addition, the upper-tropospheric jet migrates equatorward by $\sim$10° and strengthens by 20--30\%, particularly at the evening terminator.

More specifically, we find that at 100~hPa, the Baseline simulation shows a broad east–west overturning flow with weak stationary perturbations confined to mid-latitudes. The introduction of steep substellar mountains substantially modifies this pattern: the flow develops a pair of standing Rossby gyres on the dayside and a downstream ridge in the nightside mid-latitudes.  Specifically, the Steep Uplift case organizes the 100~hPa flow into a coherent, wave-locked export regime that preferentially accelerates the upper-tropospheric westerlies toward the evening terminator, consistent with an equatorward-shifted and strengthened jet (by $\sim 10^{\circ}$ and $20$--$30\%$, respectively).
In the Steep case, the  wave-jet pathway supports downstream moisture export and cloud/longwave feedbacks, thereby sustaining nightside warming and open-ocean maintenance.

In contrast, the Plateau case also breaks zonal symmetry but yields a slightly more fragmented stationary-wave pattern that is less effectively phase-aligned with the terminator export pathway, limiting the efficiency of day-night communication relative to Steep case.
The Baseline remains closest to the canonical tidally locked regime, more zonally symmetric, with weaker stationary structure and high-latitude organization, implying that cross-terminator transport is comparatively limited without the effects of mechanical waves.
Lastly, the warm region (red $273~\mathrm{K}$ contour) is no longer a smooth radiative isotherm in the topographic cases, but becomes dynamically sculpted by the downstream stationary ridge, indicating that advection and moist processes allow for the observed above-freezing conditions well into the antistellar hemisphere.

\subsection{Cross‐terminator Dynamics and Stationary‐Eddy Diagnostics}

\begin{figure*}[t!]
    \centering
    \includegraphics[width=\textwidth]{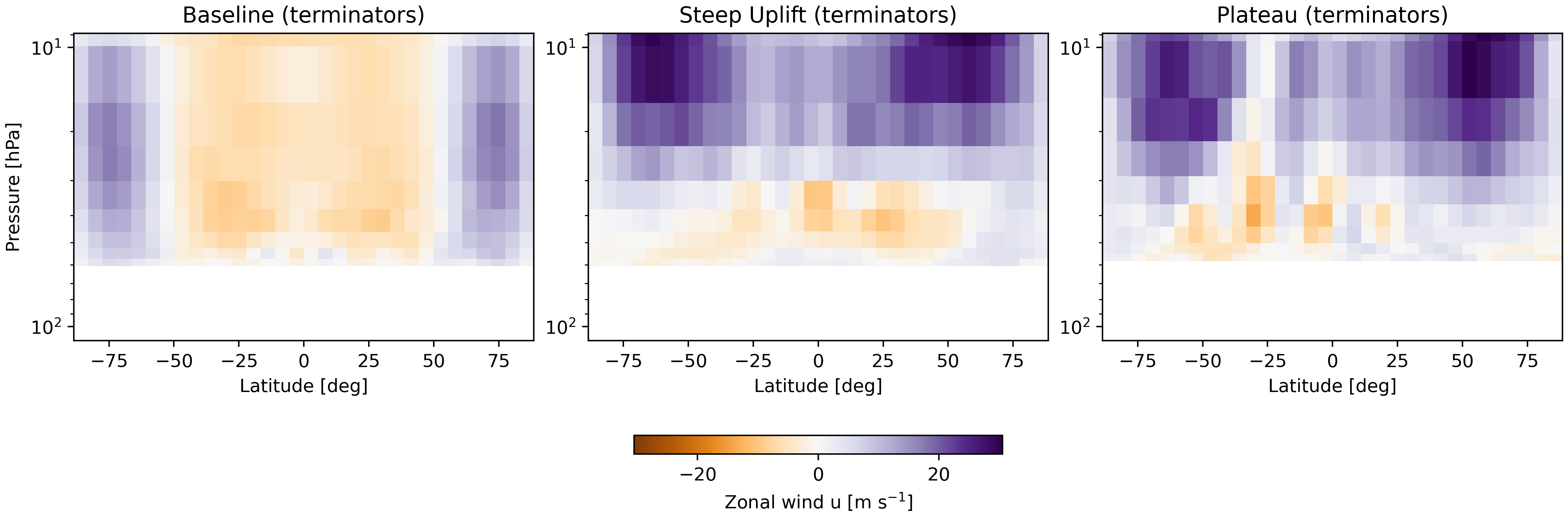}
    \caption{\textbf{Latitude–pressure cross-sections of zonal wind at the day–night terminators.}
    Each panel shows $u$ averaged over $\pm 10^\circ$ around the morning and evening terminators.
    The Steep Uplift case strengthens and deepens the upper-level westerlies and broadens the jet core, whereas the Plateau case exhibits just a slightly weaker structure. 
    Color scale is symmetric about zero.}
    \label{fig:u_terminator_sections}
\end{figure*}

Figure~\ref{fig:u_terminator_sections} shows the  zonal wind\footnote{We exclude $P>60$ hPa to focus on lower pressure dynamics where the large-scale dynamical response is more prominent. Although topography acts at the surface, it excites vertically propagating stationary waves that modify the equatorial jet aloft.} cross-sections at the day–night terminators. 
In the Baseline case, thermally driven zonal flow aloft through direct circulation is much weaker compared to the eddy driven flow, consistent with the canonical hot-spot offset pattern of tidally locked climates.
When steep mountains are introduced, the upper-level westerlies intensify by roughly 20~m\,s$^{-1}$ and extend deeper into the troposphere (while the lower-level easterlies weaken, not shown). 
This vertical coupling arises from enhanced momentum mixing driven by orographically forced stationary waves.
In contrast, the Baseline configuration displays a shallower and weaker equatorial jet structure.
Note that, in the Baseline simulation, the high-latitude circulation is dominated by zonally symmetric polar vortices. In contrast, the introduction of steep topography disrupts this symmetry, generating a pair of standing Rossby gyres in the mid-latitudes.

Previous work has shown that stronger westerlies observed at the evening terminator relative to the morning terminator arise from the phase-lagged thermal effects and ongoing wave-mean flow acceleration on the dayside (see e.g., \citealt{showman2013}).
The introduction of surface topography further amplifies this asymmetry.
While the substellar highlands act as a mechanical barrier that locally decelerates the low-level easterly flow approaching from the morning terminator, this momentum loss is not simply dissipated.
Instead, the interaction enhances stationary wave generation and vertical momentum transport, effectively transferring momentum from the surface to the upper troposphere.
This process strengthens the longitudinal advection of mass and momentum aloft, thereby increasing the efficiency of day-to-night transport despite the local topographic drag.

\begin{figure*}[t!]
    \centering
    \includegraphics[width=\textwidth]{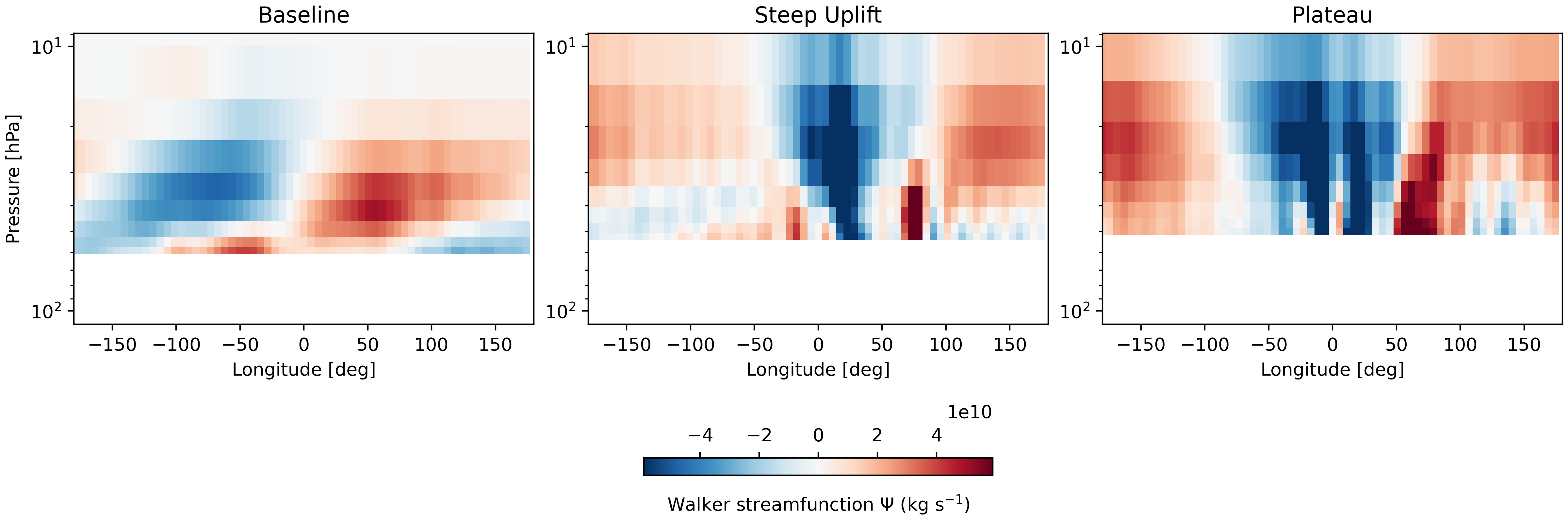}
\caption{\textbf{Equatorial Walker streamfunction $\Psi$ in longitude--pressure space.} 
Shown are the zonally overturning circulation patterns for the 
(\textit{left}) Baseline, (\textit{center}) Steep Uplift, and (\textit{right}) Plateau simulations. 
The Baseline case is characterized by broad, smoothly varying overturning spanning much of the longitudinal domain. 
In the Steep Uplift configuration, the circulation breaks into multiple vertically stacked cells with strong longitudinal modulation, consistent with vigorous stationary wave activity and enhanced three-dimensional mixing. 
The Plateau case lies between these extremes, maintaining a primary central overturning structure with superimposed secondary cells.}
\label{fig:walker}
\end{figure*}

The Baseline simulation exhibits two large, coherent Walker-type overturning cells that span most of the longitudinal domain, indicative of a relatively laminar day–night circulation (Figure~\ref{fig:walker}). In contrast, the steep-mountain configuration fragments this circulation into a series of smaller-scale, vertically segmented overturning cells. This transition reflects the dominance of stationary wave–driven and eddy-mediated transport in the presence of strong topographic forcing. Despite the loss of a single planetary-scale Walker cell, the steep-mountain case exhibits enhanced vertical mixing and repeated zonal export of moist static energy, which together promote efficient thermal redistribution and sustained nightside warming. The Plateau configuration occupies an intermediate regime: it retains a dominant central overturning cell characteristic of planetary-scale organization, but also hosts multiple embedded sub-cells and pronounced vertical segmentation near the substellar longitude.

The equatorial Walker circulation (Figure \ref{fig:walker}) reveals how topography dramatically changes the day-night coupling.
The Baseline (left) features a single, laminar overturning cell spanning the domain.
While the Plateau case is much closer to the Steep Uplift case, these two cases differ in a few important ways: First, at the Morning Terminator ($-90^{\circ}$): The Plateau case (right) exhibits a massive, coherent overturning cell (the continuous red region from $-150^{\circ}$ to $-50^{\circ}$). This "reverse-flow" suggests a laminar recirculation that  isolates the nightside.
In contrast, the Steep Uplift (center) shows a distinct counter-rotating eddy (blue region at $\sim -100^{\circ}$), indicating that the sharp topography blocks and disrupts the upstream flow. At the Evening Terminator ($+90^{\circ}$) however, the situation reverses. The Steep Uplift establishes a coherent injection cell (continuous red block) that bridges the terminator, whereas the Plateau circulation becomes fragmented (alternating cells).

The location of up- and downwelling are also modified by topography. In the Baseline case, ascent is broadly distributed and centered near the substellar longitude, with diffuse subsidence across the nightside. Introducing steep substellar relief concentrates upwelling over the mountainous region and displaces compensating downwelling eastward and deeper into the nightside, producing multiple vertically stacked overturning cells. This indicates that while broad uplift alone partially preserves large-scale circulation structure, it is already sufficient to induce topographically driven fragmentation and a transition toward wave-modulated transport.


\begin{figure*}[t!]
    \centering
    \includegraphics[width=\textwidth]{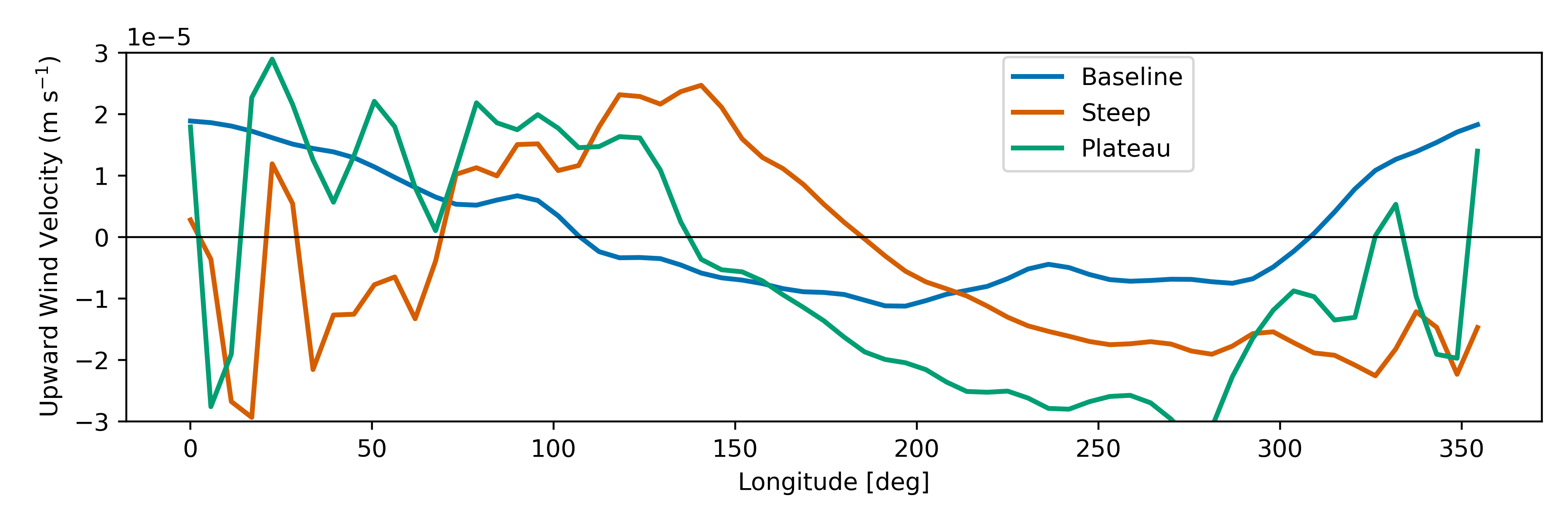}
    \caption{ Mass‑weighted vertical velocity at 200 hPa for the three surface configurations.  Compared to the Baseline case, both Steep and Plateau topographies shift ascent downstream, enhancing vertical motion on the nightside. The Steep case exhibits a sharper, more localized updraft near 120°-140° longitude, aligning with the evening terminator jet, while the Plateau case shows a broader region of enhanced ascent centered closer to the dayside. The Baseline configuration, in contrast, exhibits weaker and more symmetric vertical motion across the terminator.}
    \label{fig:wap}
\end{figure*}

The zonal variation of vertical velocity (Figure \ref{fig:wap}) reveals how topography changes the location of convective injection.
 In the Baseline simulation (blue),  vertical motion is strongest at the substellar point, mostly positive on the dayside and negative on the nightside, and varies smoothly with longitude.
Both topographic cases shift significant vertical motion downstream toward the eastern terminator ($100^{\circ}$--$140^{\circ}$); however, their structural distribution differs.
The Plateau case (green) is dominated by a massive, localized updraft deep on the dayside ($\sim 30^{\circ}$), with the terminator region appearing as a secondary feature.
In contrast, the Steep Uplift configuration (orange) concentrates its primary vertical velocity maximum on the nightside ($120^{\circ}$--$140^{\circ}$).
This terminator-aligned vertical transport effectively ``ventilates'' the dayside, injecting moisture into the westerly jets just as they cross onto the nightside, explaining the enhanced specific humidity relative to the Plateau case.

As the upward flow of the Steep case maximizes in the terminator region, this enhances fraction of convectively lofted moisture that is immediately entrained into the westerly jets at the moment they cross onto the nightside. In this geometry, vertical motion acts as a dynamical gate for nightside greenhouse effect: enhanced terminator injection increases nightside specific humidity and cloud formation, strengthens longwave optical depth, and  reduces effective radiative cooling on the dark hemisphere, which is consistent with the onset of nightside warming. Quantitatively, the Steep Uplift configuration exhibits a substantially larger upward mass flux (roughly a factor of two relative to Plateau at 200 hPa), indicating a stronger cross-terminator overturning and more efficient day–night exchange. 

We quantify the impact of topography on hemispheric energy exchange using two complementary diagnostics computed from time‐mean model output.

\textbf{(1) Cross‐terminator import of moist static energy (MSE).}  
From e..g,  \citet{pierrehumbert2010text}, the moist static energy per unit mass is
\begin{equation}
m = c_p T + L_v q ,
\end{equation}
where $c_p$ is the specific heat of air, $L_v$ the latent heat of vaporization, $T$ temperature, and $q$ specific humidity.  
At each grid point, the zonal anomaly $m' = m - \langle m \rangle_\lambda$ is multiplied by the zonal wind $u$ to yield the instantaneous longitudinal flux $u\,m'$.  
Vertically integrating and averaging within $\pm\Delta\lambda$ of each terminator gives the day–night directed flux
\begin{equation}
F_{\mathrm{night}}(\phi)
  = \max(\langle u\,m' \rangle_{\mathrm{W}},0)
  + \max(-\langle u\,m' \rangle_{\mathrm{E}},0) ,
\end{equation}
where $\langle\cdot\rangle_{\mathrm{W,E}}$ denote averages near the west ($90^\circ$) and east ($270^\circ$) terminators.  The sign change accounts for  eastward flow at the evening terminator (westerlies, or positive) and westwards flows over the morning terminator (easterlies, or negative).
The global nightside energy import is
\begin{equation}
P_{\mathrm{night}}
= 2\pi a \!\int_{-\pi/2}^{\pi/2} \! F_{\mathrm{night}}(\phi)\cos\phi\,\mathrm{d}\phi ,
\end{equation}
with $a$ the planetary radius.  $P_{\mathrm{night}}$ is reported in petawatts (PW).  
Results are insensitive to $\Delta\lambda$ between $5^\circ$ and $15^\circ$.

\textbf{(2) Meridional stationary‐eddy MSE transport.}  
To quantify how topography‐locked stationary waves redistribute energy, we diagnose the meridional flux of moist static energy (MSE) associated with stationary eddies \citep{hill2015}:
\begin{equation}
F_{\mathrm{MSE}}^{\mathrm{stat}}(\phi)
= \int \frac{1}{g}
\Big[c_p \langle v^* T^* \rangle_\lambda
   + L_v \langle v^* q^* \rangle_\lambda \Big]
\,\mathrm{d}p ,
\end{equation}
where $v$ is the meridional wind, $T$ temperature, $q$ specific humidity, and $X^*=X-\langle X\rangle_\lambda$ denotes the stationary anomaly (longitude‐fixed deviation).  
The two additive terms represent dry‐static and latent components,
\begin{equation}
F_H^{\mathrm{stat}} \propto \langle v^* T^* \rangle_\lambda, \qquad
F_Q^{\mathrm{stat}} \propto \langle v^* q^* \rangle_\lambda ,
\end{equation}
and their sum gives the total stationary‐eddy MSE transport.  
All quantities are time‐mean fields, vertically integrated in pressure coordinates ($\mathrm{d}p/g$), so that $F_{\mathrm{MSE}}^{\mathrm{stat}}(\phi)$ measures the net north–south energy flux by stationary waves at each latitude.

\begin{figure}[t]
\centering
\includegraphics[width=0.9\linewidth]{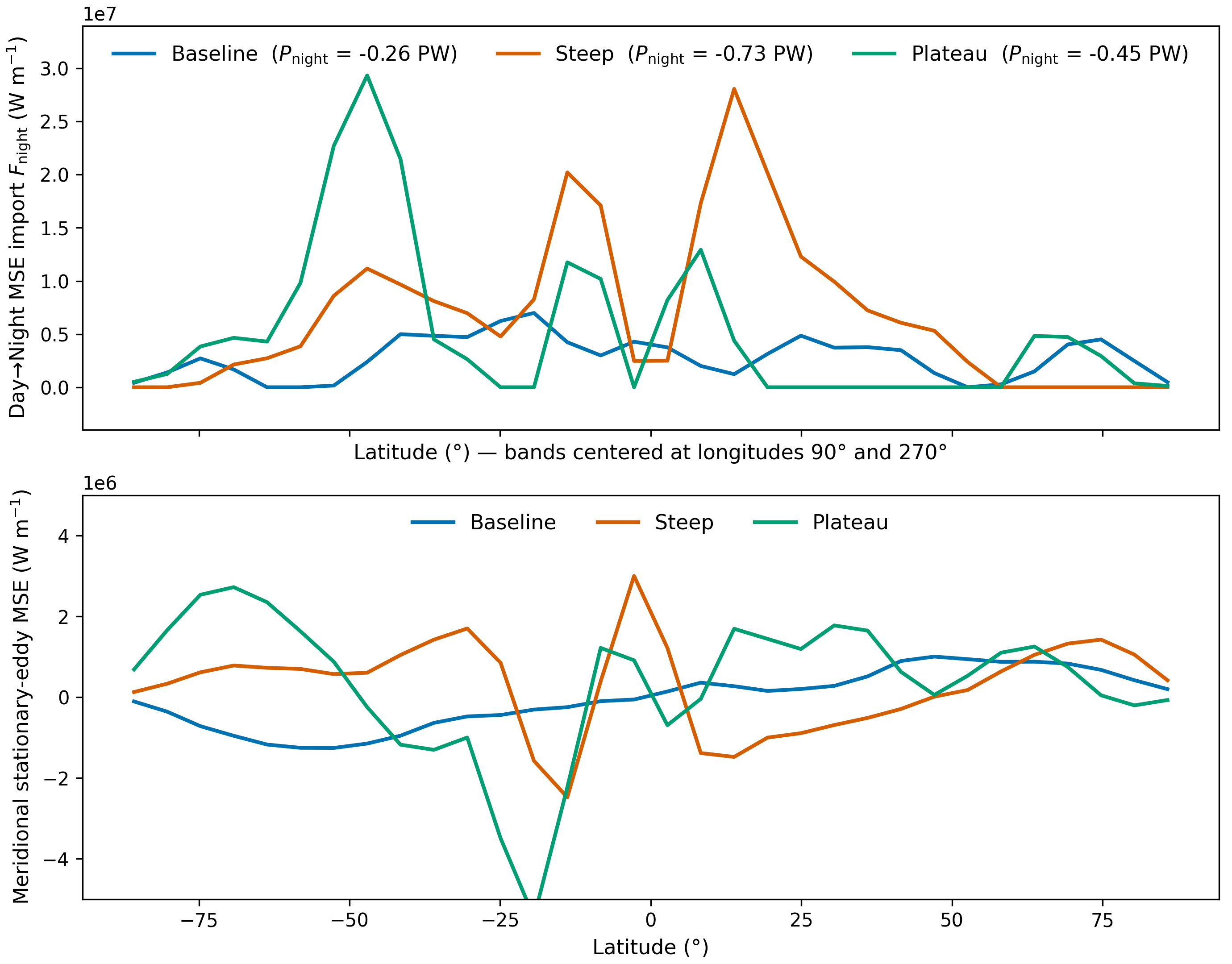}
\caption{
Day--night energy exchange and its stationary--eddy mechanism. 
\textbf{Top:} Cross--terminator import of moist static energy (MSE) as a function of latitude, 
computed from stationary--eddy fluxes of $u\,m'$ and averaged in $\pm\,\mathrm{90}^{\circ}$ 
longitude bands centered at the west and east terminators (longitudes $90^{\circ}$ and $270^{\circ}$). 
The legend lists the global nightside power $P_{\mathrm{night}}$. 
\textbf{Bottom:} Vertically integrated meridional stationary--eddy MSE transport, 
$c_p [v^*T^*] + L_v [v^*q^*]$. 
Both Steep and Plateau topographies enhance cross-terminator MSE import at mid-latitudes relative to Baseline, with Steep producing a sharper peak and Plateau a broader enhancement.
}
\label{fig:eddy}
\end{figure}

Using the above analyses, we find that topographic amplification of the stationary‐wave field substantially modifies how energy is exchanged between the day and nightside hemispheres.  Figure~\ref{fig:eddy} (top) shows the vertically integrated flux of moist static energy (MSE) across the terminators\footnote{Note that the meridional transport peaks near the the substellar ridge where the flow is diverted poleward, while the zonal transport peaks in the mid-latitudes where the flow, now laden with moisture, is injected across the terminator by the jets. The meridional transport effectively 'feeds' the zonal transport.}, computed from stationary–eddy components of $u\,m'$.  The steep case exhibits the largest hemispheric‐mean import of MSE to the nightside ($P_{\mathrm{night}}\!\approx\!-0.7$~PW), roughly three times greater than in the Baseline simulation.  This enhancement arises from stronger zonal asymmetries in the thermal and velocity fields near mid–latitudes, where the day–to–night eddy fluxes are most efficient.  The Plateau case also breaks the zonal symmetry of the flow, exhibiting significant longitudinal variability and wave generation compared to the laminar Baseline. However, its circulation structure differs from the Steep Uplift case: rather than forming a coherent, high-amplitude stationary wave train that aligns with the evening terminator, the Plateau forcing results in a  peak closer to the morning terminator.  This suggest that even small differences in the specific topographic gradient can lead to potentially large atmospheric dynamical consequences.

We note that the Plateau transport exhibits a strong hemispheric bias, peaking near $50^{\circ}$. This result is an artifact of the stochastic land generation, which concentrated the highest elevations in the southern hemisphere for this realization (see Figure~\ref{fig:topography}). This spatial correlation acts as a natural control, confirming that the stationary-wave forcing and resulting energy transport are locally locked to the specific placement of topographic landmass.

The associated meridional stationary-eddy transports  help explain the underlying mechanism (Fig.~\ref{fig:eddy}, bottom).  Both the steep and Plateau cases generate pronounced $c_p[v^*T^*]$ and $L_v[v^*q^*]$ enhancements that coincides with the latitudinal bands of enhanced day–night exchange, consistent with the excitation of quasi‐stationary wave modes by orographic forcing.  These eddies channel enthalpy and latent energy away from the substellar ridge and into the antistellar hemisphere, sustaining nightside warming even under limited direct circulation.  The Baseline case, lacking such strong stationary disturbances, shows minimal eddy transport and therefore reduced hemispheric heat redistribution. These results demonstrate that steep orography can enhance planetary‐scale stationary waves that act as conduits for moist energy flux across the terminators.

The combined diagnostics demonstrate that surface relief, whether steep or broad, acts as a first-order control on planetary climate by exciting stationary waves that modify vertical energy exchange (see e.g., \citealt{joseph2004}).
 Both topographic configurations substantially modifies the circulation relative to the Baseline, driving a transition away from the canonical ``eyeball'' regime.
Broad, smoothed relief (Plateau) is sufficient to break the hemispheric symmetry and drive an intermediate climate state, while sharper gradients (Steep) further amplify the feedback loop to sustain a fully ventilated nightside.

In terms of observational implications, we find that surface topography imprints a measurable signature on disk-integrated thermal phase curves (Figure~\ref{fig:phase}). The baseline case exhibits the largest phase-curve amplitude ($\Delta T_b = 1.82\%$) with a thermal maximum occurring near orbital phase $\sim81^\circ$, consistent with a strongly localized dayside hotspot. In contrast, enhanced topographic forcing substantially modifies the thermal phase curve, reducing the amplitude (to $\Delta T_{b}=0.89\%$ for the Steep Uplift case and $1.10\%$ for the Plateau case) and shifting the thermal peak westward to $\sim290^{\circ}$ and $\sim273^{\circ}$, respectively. Unlike the Baseline case, where the hotspot is advected eastward by the equatorial jet, the topographic cases exhibit a peak brightness temperature located upstream of the substellar point. This westward shift is a radiative signature of the stationary-wave circulation: the orographic forcing channels moisture to the east, generating a thick, optically dense cloud deck downstream of the substellar highlands. This cloud blanket suppresses outgoing longwave radiation on the eastern hemisphere, forcing the region of peak thermal emission to migrate to the clearer, drier western hemisphere. These signatures (that is, reduced amplitude and westward hotspot offset) suggest that the competition between jet-driven advection and topography-locked cloud formation could be an observable diagnostic of surface-atmosphere coupling on synchronously-rotating terrestrial exoplanets.

\newpage
\section{Discussion}

The results presented here show that the geometric distribution of relief can act as a parimary control on the climate state of synchronously rotating planets.   
In our experiments, localized mountains generate stationary planetary waves that enhance vertical mixing, strengthen cross‐hemispheric jets, and sustain nightside humidity.  
This behavior is analogous to the role of the Tibetan Plateau or Andes in Earth's circulation \citep{insel2010,huang2023}, yet under a very different radiative geometry: rather than shaping seasonal monsoons, the orography on a tidally locked planet modulates the continuous day–night overturning cell.

\begin{figure*}[t!]
    \centering
    \includegraphics[width=0.9\textwidth]{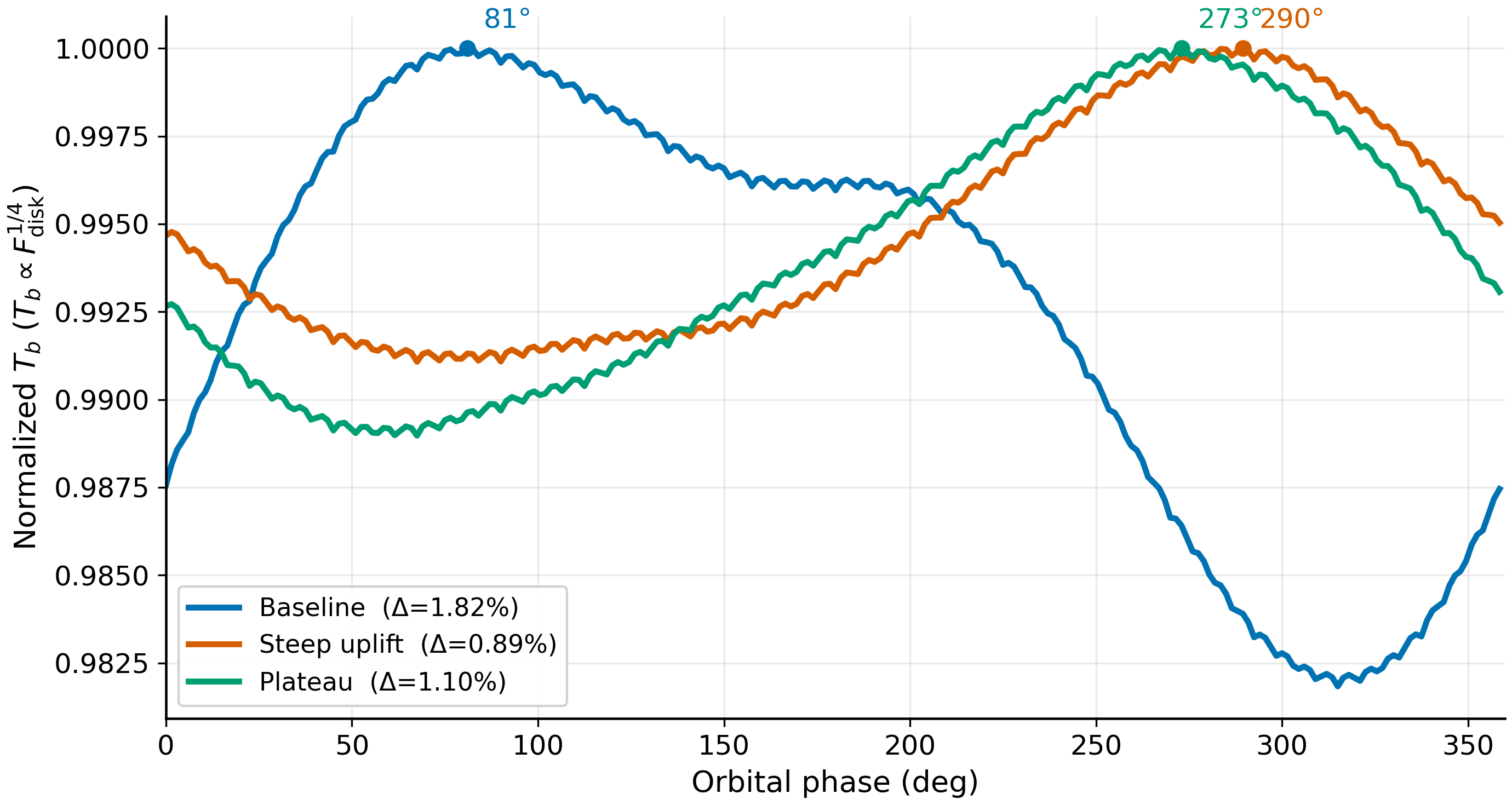}
    \caption{Normalized brightness temperature (bolometric) phase curves inferred from disk-integrated TOA longwave emission for the baseline, steep-uplift, and plateau cases. Peak emission occurs at $\sim81^\circ$, $\sim290^\circ$, and $\sim273^\circ$, respectively, with corresponding phase-curve amplitudes of $1.82\%$, $0.89\%$, and $1.10\%$. Enhanced topography both reduces the phase-curve amplitude and displaces the thermal hotspot, indicating more efficient longitudinal heat redistribution associated with topography-driven circulation and cloud feedbacks. }
    \label{fig:phase}
\end{figure*}

If a planet’s surface relief alters the efficiency of day–night energy transport by tens of percent, then the   outer edge of the liquid water habitable zone \citep{KastingEt1993Icarus,KastingEt2015ApJL} can shift. 
For planets near the threshold between eyeball” and temperate–nightside states, substellar highlands could delay global glaciation  or lower the critical background pressure for deglaciation by a few factors.  
Conversely, smooth or Plateau‐like topographies, even when equally elevated, weakens the dynamical coupling that maintains a warm nightside,   narrowing the stable temperate zone.  
Similar in spirit to the studies of land/desert planets \citep{abe2011,honing2023}, our results suggest that surface geometry should be treated as an intrinsic dimension of the habitable parameter space, comparable in importance to atmospheric mass or rotation rate.

The topographic control on stationary‐wave amplitude and cloud asymmetry may have observational consequences for next‐generation flagships.
The distinctions between differnet topographies as seen in Figure~\ref{fig:phase} may be detectable for nearby systems like TRAPPIST-1e or LHS 1140 b with future flagship mid-infrared interferometers. For JWST however, the climate effects on topography are unlikely to be observationally constrained for M-dwarf terrestrial planets in the near-term. This represents an additional, often overlooked source of uncertainty in the interpretation of atmospheric and surface observations (see also \citealt{macdonald2022}).

While our simulations capture the first‐order dynamical impacts of relief, there are still several simplications.
The model employs a fixed land albedo and omits topographically induced snow feedbacks, which could amplify contrasts between elevated and low‐lying regions.  
Ocean dynamics and horizontal heat transport are likewise neglected; including them would likely reduce the magnitude of the simulated day–night and pole–equator thermal contrasts \citep{YangEt2019ApJ,zhao2021}.  
Our dry‐continent setup also neglects runoff and latent‐heat buffering from inland water bodies that might further moderate temperature gradients.  
Future work coupling \texttt{ExoPlaSim} with  ocean, land–hydrology, and radiative–convective schemes will help determine whether the feedbacks identified here persist in more realistic settings, which are recently shown to be important \citep{di2025}. 


In the steep-uplift and plateau configurations considered here, large landmasses and uplifted continental interiors would act as barriers to zonal ocean flow, limiting the development of basin-scale currents capable of compensating atmospheric heat transport. As a result, ocean dynamics in such configurations would likely be confined to regional gyres and shallow mixed-layer exchange, reducing their ability to suppress the atmospheric contrasts driven by topographic forcing. This suggests that the atmospheric mechanisms identified in this study, stationary-wave excitation, vertical mixing, and cloud-mediated radiative feedbacks, would remain the dominant regulators of the climate state even in the presence of an active ocean

 An important avenue for future work is to examine how the surface relief explored here may influence atmospheric photochemistry through its impact on transport and cloud structure, as previous exoplanet 3D models have shown that the inclusion of interactive chemistry can result in departures from 1D models (e.g., \citealt{braam2025,ChenEt2019ApJ,luo2023,cooke2023}). The Steep Uplift configurations identified in this study suggest a regime in which orographically forced ascent and stationary-wave activity could enhance vertical mixing across the upper troposphere and lower stratosphere, potentially increasing the upward flux of photochemical precursors such as O$_2$, H$_2$O, and HO$_{\rm x}$ source species into regions of strong stellar UV irradiation. In such a regime, wave-driven longitudinal transport could also redistribute photochemically produced species across the terminator, reducing dayside confinement and increasing nightside column abundances. For species such as ozone, whose production depends sensitively on both UV exposure and vertical supply of oxygen, this raises the possibility that topography may imprint a strong hemispheric and vertical structure on both abundance and radiative impact. Enhanced cloud formation near the terminator and on the nightside, as seen here, could further modulate this  by attenuating UV flux while amplifying longwave radiative feedbacks.

Spatially varying topography thus provides a pathway for breaking hemispheric symmetry and creating long‐lived regional climates, much as continents and orography do on Earth.  

\section{Conclusions}

This study demonstrates that the spatial organization of topography can exert a strong influence on the climate of tidally locked terrestrial planets.  
We conduct climate modeling of M-dwarf planets using Baseline, Steep Uplift, and Plateau landmasses under a range of incident stellar fluxes and atmospheric gas pressures. We find that localized substellar orography excites stationary planetary waves that strengthen day–night circulation, deepen cross-terminator jets, and maintain a ventilated, cloud-blanketed nightside, and ultimately lower the critical thresholds required for global planetary deglaciation. Topography, therefore, represents an overlooked dimension in the climate phase space of synchronously-rotating worlds. 

The key findings are summarized as follows:

\begin{itemize}
    \item  While the flat Baseline features symmetric vortices and a laminar Walker cell, steep orography acts as a mechanical barrier that decelerates low-level easterlies. This excites vertically propagating stationary waves, replacing symmetric vortices with standing Rossby gyres. Consequently, the upper-tropospheric jet migrates equatorward by ${\sim}$10\%
  and strengthens by 20-30\%, accelerating westerlies toward the evening terminator.

\item  The terminator-aligned ascent acts as a dynamical gate, driving a cross-terminator moist static energy import to the nightside of $\sim0.73 \text{ PW}$, which is roughly three times greater than on a flat planet. 
This massive injection of energy and moisture sustains an optically thick nightside cloud deck that insulates the surface against radiative cooling.
Consequently, this lowers the threshold for global deglaciation, enabling a 100\% ice-free ocean at transitional parameters (e.g., 6.0 bar) where a flat planet retains a massive nightside ice shell and only 39\% open water. 

    \item These circulation shifts manifest as distinct thermal phase curve anomalies, characterized by a westward hotspot shift of over $200^\circ$ and dampened thermal amplitude due to the suppression of outgoing longwave radiation by downstream clouds.
\end{itemize}

Future global circulation and radiative–convective models that include realistic surface relief, coupled hydrology, and variable rotation rate will be critical for assessing the prevalence of these feedbacks across the M-dwarf planet population.  
Such efforts will help gain a more complete understanding of how geology, dynamics, and radiative transfer  determine the long-lived climates on M-dwarf exoplanets.

\section*{Data Availability}

The analysis code supporting the findings of this study 
is available on https://doi.org/10.5281/zenodo.18394195, and the full GCM dataset is available upon request.

\acknowledgements
The project was partially funded by the European Union (ERC, EASE, 101123041). Views and opinions expressed are however those of the author(s) only and do not necessarily reflect those of the European Union or the European Research Council Executive Agency. Neither the European Union nor the granting authority can be held responsible for them. H.C. acknowledges the College of Engineering and Science for research support.

\software{ExoPlaSim \citep{Paradise+22}; Matplotlib \citep{Hunter2007}}

\bibliographystyle{apj}

\end{document}